\documentclass{article}

\usepackage{arxiv}
\usepackage[utf8]{inputenc} 
\usepackage[T1]{fontenc}    
\usepackage{hyperref}       
\usepackage{url}            
\usepackage{booktabs}       
\usepackage{amsfonts}       
\usepackage{nicefrac}       
\usepackage{microtype}      
\usepackage{graphicx}

\title{Revolutionising Distance Learning: A Comparative Study of Learning Progress with AI-Driven Tutoring}

\author{
 Moritz Möller\thanks{Corresponding author.} \\
  IU International University of Applied Sciences\\
  \texttt{moritz.moeller@iu.org} \\
  \And
 Gargi Nirmal \\
  IU International University of Applied Sciences\\
  \texttt{gargi.nirmal@iu.org} \\
  \And
 Dario Fabietti \\
  IU International University of Applied Sciences\\
  \texttt{dario.fabietti@iu.org} \\
  \And
 Quintus Stierstorfer \\
  IU International University of Applied Sciences\\
  \texttt{quintus.stierstorfer@iu.org} \\
  \And
 Mark Zakhvatkin \\
  IU International University of Applied Sciences\\
  \texttt{mark.zakhvatkin@iu.org} \\
  \And
 Holger Sommerfeld \\
  IU International University of Applied Sciences\\
  \texttt{holger.sommerfeldt@iu.org} \\
  \And
 Sven Schütt \\
  IU International University of Applied Sciences\\
  \texttt{sven.schuett@iu.org} \\
}

\begin{document}
\maketitle
\begin{abstract}
Generative AI is expected to have a vast, positive impact on education; however, at present, this potential has not yet been demonstrated at scale at university level. In this study, we present first evidence that generative AI can increase the speed of learning substantially in university students. We tested whether using the AI-powered teaching assistant Syntea affected the speed of learning of hundreds of distance learning students across more than 40 courses at the IU International University of Applied Sciences. Our analysis suggests that using Syntea reduced their study time substantially—by about 27\% on average—in the third month after the release of Syntea. Taken together, the magnitude of the effect and the scalability of the approach implicate generative AI as a key lever to significantly improve and accelerate learning by personalisation. 
\end{abstract}

\keywords{higher education \and university teaching \and artificial intelligence \and large language models}

\section{Introduction}
\label{sec:introduction}

The pursuit of a higher education degree is widely recognised as not only a catalyst for career advancement but also an opportunity for personal growth and intellectual enrichment. In recent years, access to such education has considerably broadened, for example due to the proliferation of distance learning programmes. These programmes provide students with heightened flexibility, moulding education around their lives rather than constraining their lives around education. Yet, even these more flexible learning pathways require a substantial investment of time, thereby posing a significant challenge for learners, especially working adults. Juggling professional commitments, personal responsibilities, and educational pursuits can often leave this demographic feeling overwhelmed. Subsequently, lack of time is one of the predominant reasons individuals either discontinue their studies or refrain from enrolling to begin with.

Reducing the required learning time (without compromising on the learning outcomes) is hence a key lever to reduce drop-out rates, increase enrolment rates and support students in reaching their goals. This can be achieved by using the learning time more efficiently, for example for interactive personal tutoring sessions instead of passive reading of a coursebook. Personal tutoring has been shown to be extremely effective in improving student success, with students who received it achieving better than 98\% of students in a control class on average in a seminal study \cite{bloom19842}. Unfortunately, such personal tutoring is very costly, hard to scale, and hence at present only accessible to a very small part of the student population.

Recent technological breakthroughs, particularly the advent of generative AI \cite{achiam2023gpt}, are bound to disrupt that status quo. AI-based learning assistants can now be found across educational offerings \cite{freeman2023duolingo, towers-clark2023khan, pinto2023large, liu2024teaching}. One of those learning assistants, named Syntea \cite{iu2024syntea}, is being developed at the IU International University of Applied Sciences, Germany’s largest university with over 100 000 students, which offers more than 200 different degree programs, many of those in distance learning formats. Launched in 2022, Syntea constitutes the first major initiative centred around generative AI at scale in higher education, and serves a single, central goal: to empower people across the globe to grow through the most personalised education.

Since its launch in 2022, Syntea has evolved to include various features, particularly an exam training feature which allows students to practise exam-relevant content and receive immediate feedback. This reflects the IU’s belief that personalised feedback-corrective procedures are key to improve learning progression. The feature was rolled out in more than 40 courses for thousands of students at the end of September 2023; these students are now able to use it at their own discretion.

In this study, we aim to examine the impact of Syntea (and especially her exam training feature) on learning progress and outcomes. We are interested in whether students learn faster when using the exam training feature—as laid out above, increasing study progression is of essential interest especially for distance learners. The flexible format of the IU distance learning degree programs is ideally suited for such analyses, as students can freely choose the time of their examination: online renderings of exams enable these students to write an exam within one hour at any point in time. 

\section{Results}
\label{sec:results}

To analyse the speed of learning, we focused on study progress as a metric, which we defined as average number of exams passed per student and month. To investigate this metric in relation to the release of Syntea, we ran a two-tiered analysis: first, we performed an explorative analysis to get a general picture of the relevant dynamics in section~\ref{subsec:exp}; second, we performed a more rigorous statistical analysis to test the robustness of the main effects we discovered in section~\ref{subsec:stat}. Finally, we summarise our results in section~\ref{subsec:sum}

\subsection{Explorative analysis}
\label{subsec:exp}

For the explorative analysis, we computed study progress over time for a treatment group (the group of students that have used Syntea's exam trainer feature at least once) and a control group (all other students with a distance learning program). To extract the signal for both groups, we used a sliding window average with a 60-days window, which we computed for each day in a time interval ranging from 24 months before till 3 months after the rollout of the assistant (October 2021 – December 2023). This time range was chosen to include a substantial period before the rollout to give a solid baseline, and to exclude recency effects related to delays in data processing. 

\begin{figure} 
    \centering
    \includegraphics{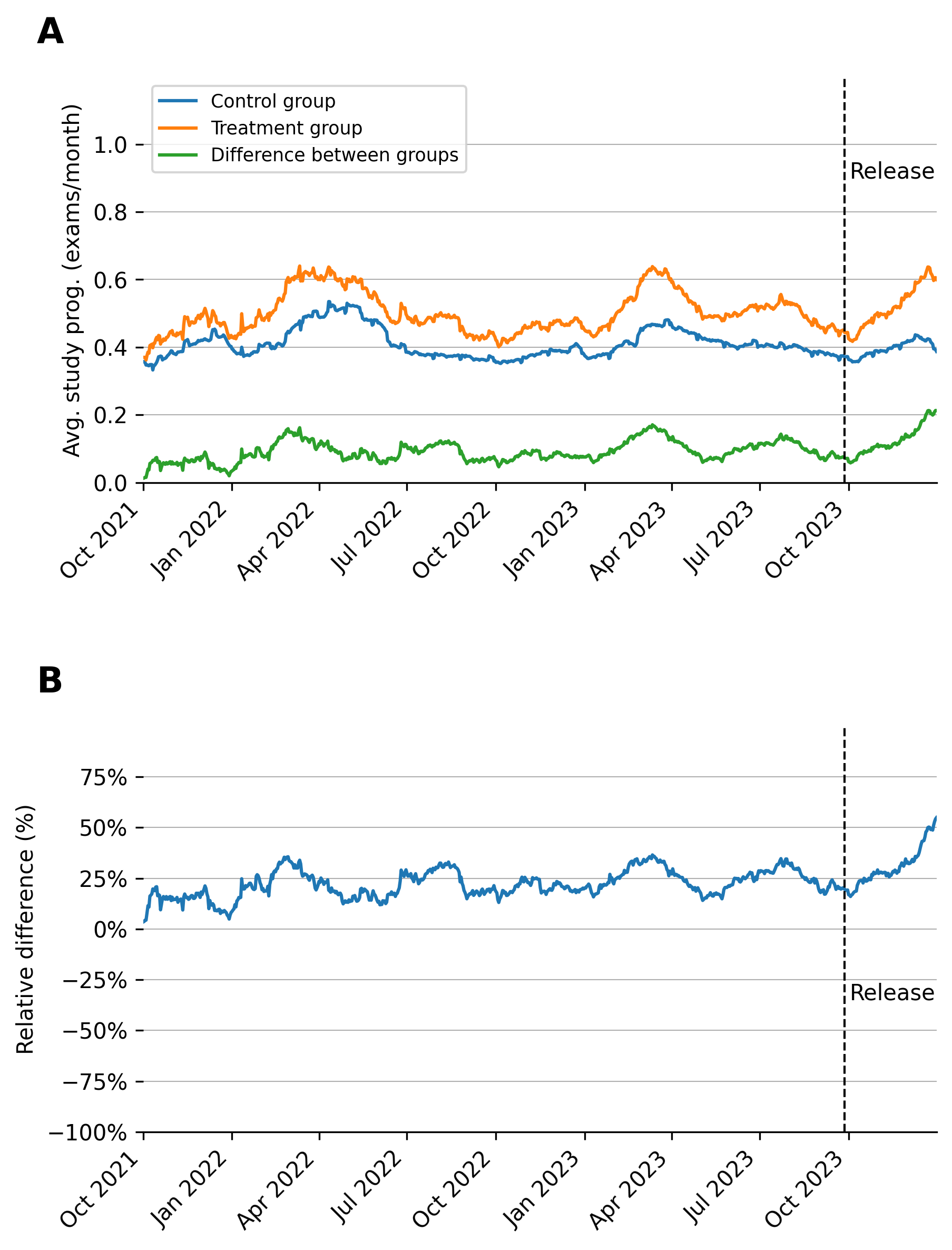}
    \caption{\textbf{Explorative analysis.} A) 60-day moving window average of study progression (y-axis) over time (x-axis) for the treatment group (orange) and the control group (blue). We also show the difference between the two groups (green). The release is marked by a dashed vertical line. B) Relative difference between treatment and control group (y-axis) over time (x-axis), with the release date marked by a vertical dashed line.}
    \label{fig:progress}
\end{figure}

The dynamics of both groups are shown in Figure~\ref{fig:progress}A, which also shows difference between both groups. In Figure~\ref{fig:progress}B, we show the relative difference, defined as the difference between treatment group and control group, divided by the control group.

We found that study progress shows some seasonality for both groups (for example, study progression appears to increase in spring and decrease in the fall). We also found that the treatment group was consistently faster than the control group, making the difference between the two groups positive throughout the investigated time range. Most importantly, we finally found that the gap between control group and treatment group widened directly after the feature was released, with an upwards trend at the time of writing. 

\begin{figure}[h]
    \centering
    \includegraphics{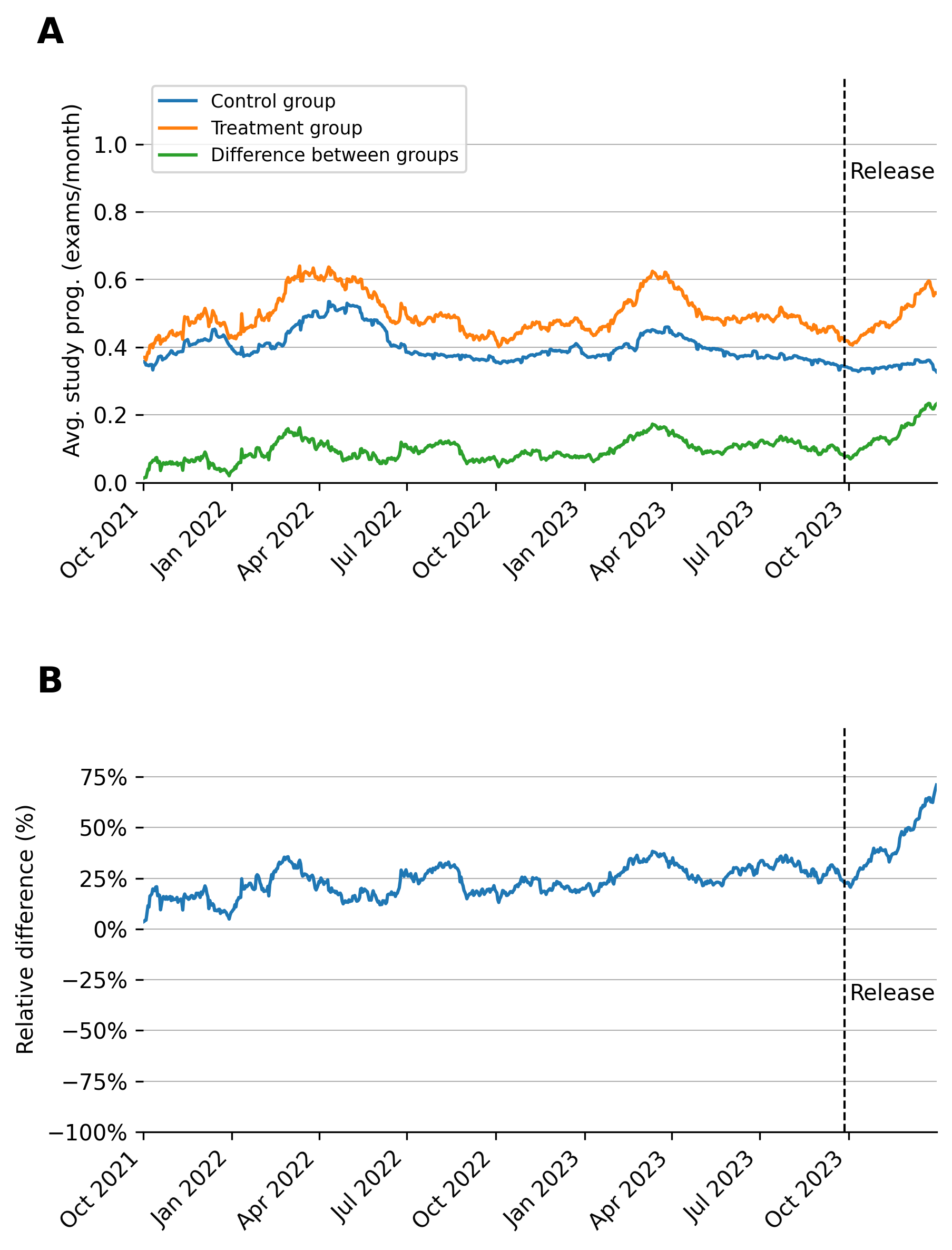}
    \caption{\textbf{Explorative analysis without new joiners.} We show the exact same data as in Figure~\ref{fig:progress} above, with one exception: we excluded all students that have started in or after January 2023, from both the control group and the treatment group. We find that the main dynamical features persist (compare Figure 1). In particular, we find that the widening gap between control and treatment group that follows the release date is present, and in fact even more pronounced with new joiners excluded.}
    \label{fig:progress_sup}
\end{figure}

To ensure that these effects are not due to a recent change in student population, we conducted a supplementary analysis where we excluded all students that joined in or after January 2023 (see Figure~\ref{fig:progress_sup} in section \ref{sec:supplementary}). We found no noticeable difference in the dynamics, which suggests that the dynamics observed are not due to a recent change in student population, but instead to a genuine change in student behaviour.

\subsection{Statistical analysis} 
\label{subsec:stat}

To determine the statistical significance and size of this effect, we first aggregated our data on month-level (i.e., we computed the number of exams for each student and month within the relevant time frame). We then computed a baseline for each month by averaging across the entire control group. This baseline then was used to adjust the measurements in the treatment group by subtracting and dividing by the baseline month-by-month, hence obtaining the relative difference in study progression per month and student between the treatment group and the control group.

We compared the average of those relative differences pre- and post-release (see Table~\ref{tab:progression}). We found that there was a positive but non-significant difference between the 24 months prior and the first month after the release (+3.8 percent-points, p = 0.58). However, there was a larger, positive and statistically significant difference between the same prior period and the second month after the release (+14.9 percent-points, p = 0.03), and an even higher difference in the third month (+45.8 percent-points, p < 0.001). Our findings suggest that the widening gap effect from the explorative analysis is not just a statistical fluctuation or artifact of the moving window averaging method, but a real feature of the underlying dynamics.

\begin{table}
\caption{\textbf{Study progression statistics.} We show averages of study progression and study duration per student (and the corresponding numbers of observations) both for the control and the treatment group for different periods of time. We further show the relative difference between the groups and how this difference changed after the release. Next, we show significance test outcomes for the pre-post change of the relative difference between the groups (t-test for two independent samples). Finally, we show the ratio of study duration between the treatment group and the control group, and the pre-post changes of this ratio.}
\label{tab:progression}
\centering
\begin{tabular}{lllll}
\toprule
&Months -24 to -1 pre &Month 0 post &Month 1 post &Month 2 post \\
&(10/21 – 09/23) &(10/23) &(11/23) &(12/23) \\
\midrule
\multicolumn{2}{l}{\textbf{Control group}} \\
\midrule
Avg. study prog. (exams/month) &0.41 &0.42 &0.40 &0.36 \\
Avg. study duration (months/exam) &2.43 &2.38 &2.49 &2.77 \\
Number of observations &1110788 &57630 &57769 &57297 \\
\midrule
\multicolumn{2}{l}{\textbf{Treatment group}} \\
\midrule
Avg. study prog. (exams/month) &0.51 &0.54 &0.56 &0.61 \\
Avg. study duration (months/exam) &1.98 &1.86 &1.79 &1.63 \\
Number of observations &13178 &960 &994 &1010 \\
\midrule
\multicolumn{2}{l}{\textbf{Relative diff. between treat. and contr.}} \\
\midrule
Avg. study prog. &24.1\% &27.9\% &39.0\% &69.9\% \\
Pre-post comparison & &+3.8\% &+14.9\% &+45.8\% \\
T-statistic & &0.55 &2.16 &5.92 \\
P-value & &0.584 &0.031 &< 0.001 \\
\midrule
\multicolumn{2}{l}{\textbf{Ratio between treat. and contr.}} \\
\midrule
Avg. study duration &80.6\% &78.2\% &71.9\% &58.9\% \\
Pre-post comparison & &-2.4\% &-8.6\% &-21.7\% \\
\bottomrule
\end{tabular}
\end{table}

\subsection{Summary of results} 
\label{subsec:sum}

Our analysis yields two key results: 

\begin{enumerate}
    \item Students that now use Syntea have always been faster than the average distance learner (24.1\% on average in the two years prior to the release).
    \item Since the release of Syntea, these Students are accelerating: their lead increased from 24.1\% before the release to 69.9\% in the third months after the release, with an upwards trend at the time of writing.
\end{enumerate}

\begin{figure}
    \centering
    \includegraphics{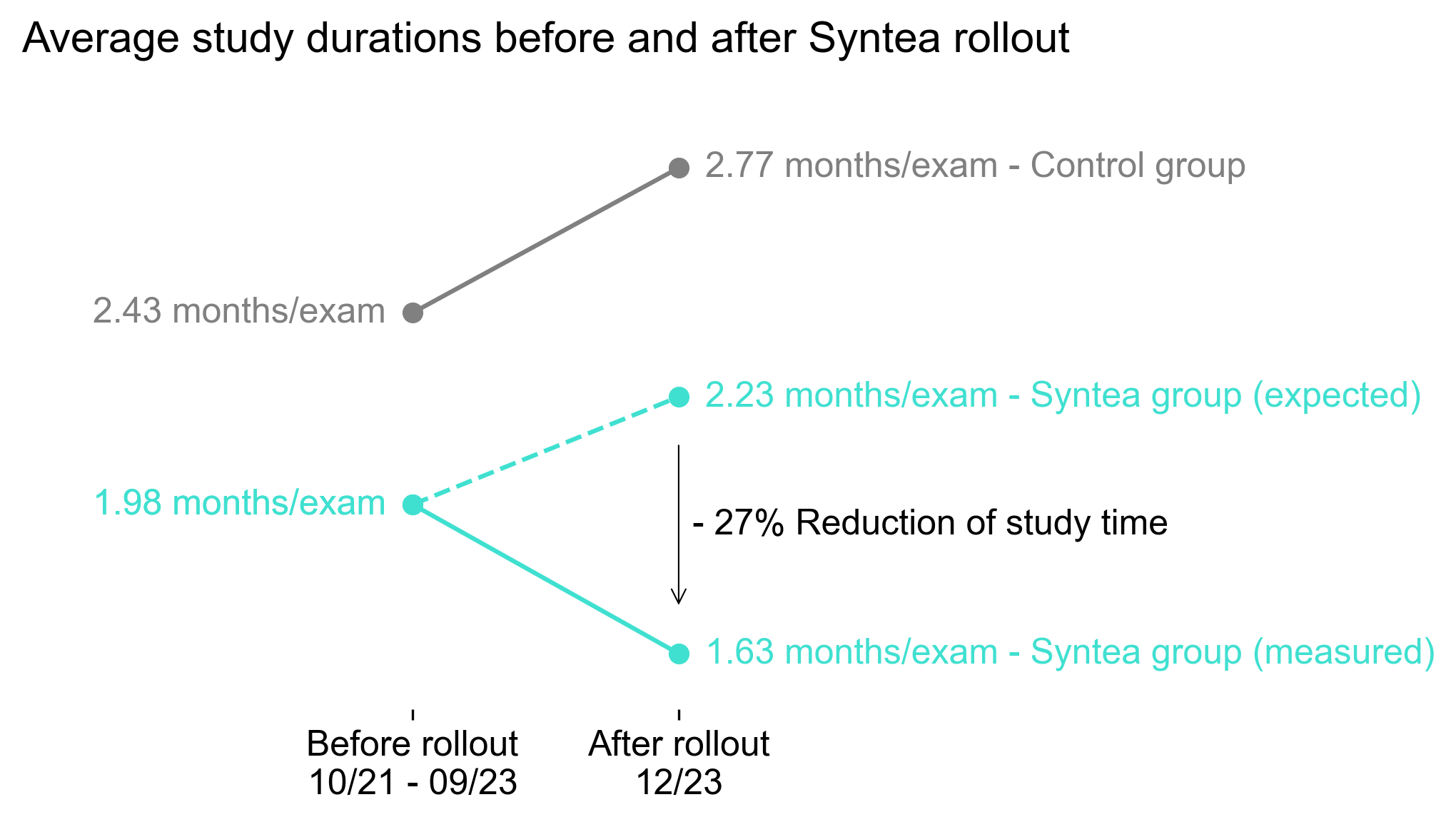}
    \caption{\textbf{Average study duration before and after Syntea rollout.} We show a summary of our main results, taken from Table~\ref{tab:progression}. We focus on the average study duration (using units of months/exam) of both the control group and the treatment group (called 'Syntea group' above), comparing the period before the release of Syntea (the months -24 to -1 prior to release, 10/21 - 09/23) with the most recent period (month 2 post release, 12/23). The expected value for the treatment group after the rollout is based on the ratio between treatment group and control group in the pre-rollout period (80.6\%), which is extrapolated to the post-rollout period (yielding a prediction of $2.77 \textrm{ months/exam}\times 80.6\% = 2.23 \textrm{ months/exam}$). The relative difference between this expected value and the measured value of 1.63 months/exam constitutes a reduction in average study time of 27\%.}
    \label{fig:progress_summ}
\end{figure}

Our results suggest that using Syntea (in this case represented by the exam training feature) has helped students to gain an additional uplift in study progress of 46 percent-points relative to the control group. Translated into study duration ratios, this corresponds to a reduction from 81\% relative to the control group to 59\%, i.e. by about 22 percent-points, or equivalently by 27\% (see Figure~\ref{fig:progress_summ} for a visualisation of this result). While there might be other explanations for the data presented above, the precise temporal alignment of the effect supports this hypothesis.

To make these results more tangible, for illustration purposes assume that an average distance learning student from the control group would take 100 days to complete a given course. Before the release of Syntea exam trainer, a typical future user of the feature would have completed the same course in 81 days (potentially due to increased motivation or other so far unknown factors). After the release, the time needed would drop to 78, 72 and finally 59 days with each month passing.

\section{Discussion}
\label{sec:discussion}

One potential criticism of our approach might concern the definition of the treatment group: it could be argued that by including all students that used Syntea (i.e. the exam trainer feature), we introduce selection bias. This argument is valid when discussing absolute numbers, such as the difference between treatment and control group at a given point in time. Indeed, we observe a significant difference between these two groups throughout the entire time frame--this is likely due to confounding variables such as motivation (a motivated student might be more likely to try new learning tools, and at the same time have a higher study progression), and not due to our feature. However, such selection bias cannot explain the differences between pre- and post-release values--the widening gap described above--our main result hence remains valid despite self-selection.

Another potential criticism might aim at natural dynamics: could the effects that we observe in the treatment group simply be part of the normal drift in the signal? Indeed, we did observe that average study progression is not stable over time but shows seasonal fluctuations. We control for this effect by focussing on the baseline-corrected study progression (relative difference between treatment and control group), instead of the raw study progression in the treatment group. Through this, we can control for natural dynamics that affect both groups (compare Figure~\ref{fig:progress}A and Figure~\ref{fig:progress}B--the seasonality effects in Figure~\ref{fig:progress}A are almost completely removed in Figure~\ref{fig:progress}B, leaving only random noise). Our statistical analysis was also performed on the baseline-corrected signal, which makes natural dynamics unlikely as an explanation.

A final criticism might not target our measurement methodology, but instead the teaching intervention itself--what if Syntea did in fact speed up students, but at the cost of worse outcomes? For example, it would be easy to just send students to the exam earlier, and hence reduce the time invested in learning--however, this would likely decrease grades substantially, and also negatively affect the probability of passing the exam. In contrast, the aim of Syntea is to accelerate learning without compromising on outcomes, as described above.

To control for such side-effects, we conducted the above pre-post analysis for these variables as well; the results can be found in the supplementary materials (Tables \ref{tab:pass} and \ref{tab:grade}). We found no significant effect in pass rates (which hover between 0.98 and 0.97), but a minor reduction in the average grade (from 80.72 out of 100 pre-release to 78.31 out of 100 in the third month after the release) in the treatment group which was not observed in the control group (the corresponding values were 80.78 out of 100 and 81.22 out of 100). While this change of the difference between groups was statistically significant, the effect size was very small (-3.4 percent points, p < 0.001) compared to the acceleration reported above (+45.8 percent-points). A change of this magnitude might not be noticeable at the level of the individual student, in particular if accompanied by a significant reduction of study time needed.

\section{Conclusion}
\label{sec:conclusion}

At the time of writing and as far as we are aware, our results above constitute the world’s first demonstration of generative AI increasing study progress in the context of university education. While the magnitude of the demonstrated effect is already substantial, we expect that this is only be the beginning—IU and other institutions will keep on building, extending, and improving AI learning companions like Syntea, and further innovations in that space will likely speed up learning more and more.

A special unique aspect of the analysis given here, as well as of Syntea overall, is that the system and its impact is not limited to a certain course, or even domain. While generative AI has been used to support higher education \cite{liu2024teaching}, the efforts that we are aware of have been specialised and were driven by single departments. In contrast, the Syntea initiative at IU encompasses all subject areas (from business to technical subjects and beyond), which allows for large N in experiments like the one reported here, as well as for large volumes of data generated through usage. Those assets ultimately enable robust, data-driven optimisation, and hence help IU to increase Syntea’s capability to accelerate learning and improve learning outcomes. 

The results presented in this study have several implications regarding future work. First, in light of the proven effectiveness of learning supported by generative AI, IU will double down on efforts to extend their AI teaching assistant and make it available to an ever-larger group of students. Second, we aim at tracking the above results further, and at gaining a better understanding of 1) how exactly the feature is used and 2) what causes the observed acceleration of learning, as well as extending the analysis to other features.

Overall, this study shows a very promising snapshot of the potential of generative AI to speed up students in their learning efforts. In addition to the deployment for university learning described above, IU is also applying Syntea in online A level programs. It stands to reason that results similar to those reported here can be achieved across many learning verticals ranging from K12 to vocational learning and corporate training. Transferring the estimated reduction of learning times by about 27\% across all learning sectors would have strong effects on all learners individually, as well as on society at large. Combined with the scalability and cost-effectiveness of the approach, AI-based teaching assistants will play a crucial role in improving and democratising education at a global scale.

\section{Acknowledgements}
\label{sec:ack}

The authors would like to thank the Syntea exam trainer team and the Syntea data analytics team for enabling the results presented above. Further, thanks to Hendrik Seelinger for valuable discussions.

\bibliographystyle{unsrt}  
\bibliography{references}

\newpage

\section{Supplementary material}
\label{sec:supplementary}

As mentioned in the main text, we conducted several additional analysis to support our main results. 

\begin{table}[h]
\caption{\textbf{Pass rate statistics.} We show averages of pass rates (and the corresponding numbers of observations) both for the control and the treatment group for different periods of time. We further show the relative difference between the groups and how this difference changed after the release. Finally, we show significance test outcomes for the pre-post change of the difference between the groups (t-test for two independent samples). The approach taken here is similar to that taken in Table~\ref{tab:progression}.}
\label{tab:pass}
\centering
\begin{tabular}{lllll}
\toprule
&Months -24 to -1 pre &Month 0 post &Month 1 post &Month 2 post \\
&(10/21 – 09/23) &(10/23) &(11/23) &(12/23) \\
\midrule
\multicolumn{2}{l}{\textbf{Control group}} \\
\midrule
Avg. pass rate ($p(\textrm{pass exam})$) &0.98 &0.98 &0.98 &0.98 \\
Number of observations &335471 &17375 &17005 &15678 \\
\midrule
\multicolumn{2}{l}{\textbf{Treatment group}} \\
\midrule
Avg. pass rate ($p(\textrm{pass exam})$) &0.98 &0.98 &0.97 &0.97 \\
Number of observations &4943 &360 &396 &442 \\
\midrule
\multicolumn{2}{l}{\textbf{Relative diff. between treat. and contr.}} \\
\midrule
Avg. pass rate &-0.0\% &-0.1\% &-0.9\% &-1.1\% \\
Pre-post comparison & & -0.1\% &-0.9\% &-1.0\% \\
T-statistic & &-0.16 &-1.42 &-1.68 \\
P-value & &0.871 &0.155 &0.093 \\
\bottomrule
\end{tabular}
\end{table}

\begin{table}[h]
\caption{\textbf{Grade statistics.} We show averages of student grades (and the corresponding numbers of observations) both for the control and the treatment group for different periods of time. We further show the relative difference between the groups and how this difference changed after the release. Finally, we show significance test outcomes for the pre-post change of the difference between the groups (t-test for two independent samples). The approach taken here is similar to that taken in Table~\ref{tab:progression}}
\label{tab:grade}
\centering
\begin{tabular}{lllll}
\toprule
&Months -24 to -1 pre &Month 0 post &Month 1 post &Month 2 post \\
&(10/21 – 09/23) &(10/23) &(11/23) &(12/23) \\
\midrule
\multicolumn{2}{l}{\textbf{Control group}} \\
\midrule
Avg. grade (out of 100) &80.78 &81.46 &81.44 &81.22 \\
Number of observations &328459 &17038 &16889 &15823 \\
\midrule
\multicolumn{2}{l}{\textbf{Treatment group}} \\
\midrule
Avg. grade (out of 100) &80.72 &81.53 &80.10 &78.31 \\
Number of observations &4949 &355 &386 &458 \\
\midrule
\multicolumn{2}{l}{\textbf{Relative diff. between treat. and contr.}} \\
\midrule
Avg. grade &-0.1\% &0.1\% &-1.7\% &-3.6\% \\
Pre-post comparison & &0.2\% &-1.5\% &-3.4\% \\
T-statistic & &0.28 &-1.88 &-4.23 \\
P-value & &0.781 &0.060 &< 0.001 \\
\bottomrule
\end{tabular}
\end{table}

\end{document}